# Influence of cumulenic chains on the vibrational and electronic properties of sp/sp$^2$ amorphous carbon.

L. Ravagnan[1], P. Piseri[1], M. Bruzzi[2], S. Miglio[2], G. Bongiorno[1], A. Baserga[3], C.S. Casari[3], A. Li Bassi[3], C. Lenardi[4], Y. Yamaguchi[5], T. Wakabayashi[6], C.E. Bottani[3], P. Milani[1,*]

[1] CIMAINA - Dipartimento di Fisica, Università di Milano, Via Celoria 16, I-20133 Milan, Italy

[2] *INFN - Dipartimento di Energetica*, Via S. Marta 3, I-50139 Florence, Italy –

[3] *NEMAS - Dipartimento di Ingegneria Nucleare, Politecnico di Milano*,

Via Ponzio 34/3, I-20133 Milan, Italy

[4] CIMAINA - Istituto di Fisiologia Generale e Chimica Biologica,

Via Trentacoste 2, I-20134 Milan, Italy

[5] Department of Mechanophysics Engineering, Osaka University,

2-1 Yamadaoka, Suita, Osaka 565-0871, Japan

[6] Department of Chemistry, School of Science and Engineering, Kinki University, Kowokae 3-4-1,

Higashi-Osaka 577-8502, Japan

## Abstract

We report the production and characterization of a form of amorphous carbon films with *sp/sp$^2$* hybridization (atomic fraction of sp hybridized species $\geq$ 20%) where the predominant sp bonding appears to be $(=C=C=)_n$ cumulene. Vibrational and electronic properties have been studied by *in situ* Raman spectroscopy and electrical conductivity measurements. Cumulenic chains are substantially stable for temperatures lower than 250 K and they influence the electrical transport properties of the *sp/sp$^2$* carbon through a self-doping mechanism by pinning the Fermi level closer to one of the mobility gap edges. Upon heating above 250 K the cumulenic species decay to form graphitic nanodomains embedded in the *sp$^2$* amorphous matrix thus reducing the activation energy of the material. This is the first example of a pure carbon system where the *sp* hybridization influences bulk properties.

*Corresponding author: pmilani@mi.infn.it

Pure carbon solids are characterized by an amazingly wide variety of structures and properties originating from the combination of $sp^2$ and $sp^3$ hybridizations [1]. Carbon atoms can also organize themselves with $sp$ hybridization and the existence of systems where $sp$ and $sp^2$ bondings coexists, originating very interesting optical and electronic properties, have been predicted [2, 3]. To date no experimental results have been provided to support these predicted properties: the extreme reactivity and the tendency of $sp$ structures to undergo cross-linking reactions to form $sp^2$ structures are major obstacles making the production and manipulation of pure carbon materials containing $sp$ hybridization very difficult [4].

Isolated $sp$ carbon species are linear structures with either alternating single and triple bonds $(-C\equiv C-)_n$, called polyynes, or with identical double bonds $(=C=C=)_n$, called cumulenes [5]. Cumulenes are expected to tend to a metallic behavior, characterized by the homogeneous distribution of the $\pi$ orbitals, as their length increases, while polyynes are expected to be semiconducing even for infinitely long chains, due to bond length alternation opening an energy gap at the edge of the Brilluoin zone [6, 7].

$Sp$ chains can be stabilized by isolation in rare gas or inorganic matrices [8, 5] or terminating the chain ends with suitable functional groups [6, 9]. Calculations on infinite carbon chains indicate that polyynes are energetically more stable than cumulenes due to Peierls distortion [10], however for finite length chains the gain in energy is too low to be significant even a T = 0 K [11]. Furthermore it has been shown that the electronic configuration of the chains is determined, regardless their length, by the group termination of the chain [6]. A conspicuous amount of experimental and theoretical results are available on the electronic structure and vibrational properties of isolated polyynes [12-14], whereas almost no experimental characterizations can be found on cumulenes.

The possibility of stabilizing $sp$ chains in cluster-assembled nanostructured carbon films [15] and inside carbon nanotubes [16, 17] has been demonstrated showing that pure carbon $sp^2$ systems containing $sp$ hybridization can be synthesized. In particular the Raman spectral features at frequencies larger than 1900 cm$^{-1}$ in *sp*-containing cluster-assembled carbon films have been interpreted as caused by the stretching modes of polyynes with a smaller contribution that has been attributed to cumulenes [15, 18].

Here we report on the production and characterization of amorphous *sp/sp$^2$* carbon films where the dominant *sp* species are cumulenes. This is obtained by supersonic cluster beam deposition on a substrate kept at a temperature of 150 K. We show that cumulenes are stable up to a temperature of roughly 210 K and that they influence the electrical transport properties of the films acting as metallic doping species. By rising the film temperature to 325 K cumulenes undergo a reorganization inducing the formation of nanometric graphitic islands in the amorphous *sp$^2$* matrix, while the amount of polyynes remains substantially constant in the investigated temperature range.

Cluster-assembled films were grown in ultra high vacuum (~ 10$^{-9}$ mbar) by depositing on glass substrates a supersonic beam of carbon clusters produced by a pulsed microplasma cluster source as described in detail in refs. [19, 20]; two parallel gold contacts were previously evaporated on the substrate (200 nm thick, 4 mm long and 0.8 mm separation). The electrical conductivity at various temperatures of films with a thickness of 400 nm (applying a constant bias of 100 V) was characterized *in situ* by a Keithley 6517A electrometer with a d.c. current sensitivity of about 0.1 pA. The *sp/sp$^2$* bond ratio evolution of the films was monitored by *in situ* Raman spectroscopy using the 532 nm line of a frequency-doubled Nd:Yag laser.

In Fig. 1A we report a typical Raman spectrum of a film deposited and kept at 150 K: it presents two prominent features in the 1200-1700 cm$^{-1}$ and 1900-2300 cm$^{-1}$ spectral regions. The D and G bands between 1200 and 1700 cm$^{-1}$ are typical of amorphous *sp$^2$* carbon and are related to vibrational modes of *sp$^2$*-hybridized sites [21]. The band between 1900 and 2300 cm$^{-1}$ (C band) is associated to stretching modes of linear *sp* hybridized carbon structures and is composed by two components (fitted to gaussian peaks) at about 2100 cm$^{-1}$ (C1) and at about 1980 cm$^{-1}$ (C2). Comparing the present spectrum with those obtained from samples deposited at room temperature (RT) [15] one observes that the overall amount of *sp* species is considerably increased as compared to the $sp^2$ component and that the C2 peak is much more intense than the C1 at odd with what has been observed for RT depositions. Keeping the substrate temperature constant at 150 K for a few days no noticeable changes in the Raman spectra were observed.

In order to understand the nature the $sp$ components and their contribution to the vibrational spectra we increased *in situ* the film temperature at a constant rate (2 K/min) up to a fixed value, we kept the temperature constant for about one hour, then we decreased it at constant rate (2 K/min) back to 150 K. We monitored the evolution of the $sp^2$ and $sp$ contributions acquiring Raman spectra at fixed time intervals during the temperature cycle: we observed irreversible changes in the Raman spectra during the temperature increase, while no changes during the cooling. In Fig. 1A we report the Raman spectra measured at the end of every temperature cycle (labeled with the maximum temperature reached during the corresponding thermal cycle): the C2 peak decreases in intensity starting from 200 K and it reduces substantially reaching 325 K, while the C1 peak remains almost unchanged. The shape of the D+G band is modified and its maximum shifts towards higher frequencies.

In Fig. 1B we plot the ratio between the integrated intensity of the C band and of the D+G band: $I_C^{rel} = I_C/I_{DG}$. This quantity can be used to resolve quantitatively the evolution of the $sp/sp^2$ bond ratio in the film and it can be defined also for the C1 ($I_{C1}^{rel} = I_{C1}/I_{DG}$) and C2 components ($I_{C2}^{rel} = I_{C2}/I_{DG}$) respectively. From $I_C^{rel}$ the absolute amount of $sp$ bonds could also be determined knowing the ratio $\alpha$ between the Raman cross sections of $sp$ and $sp^2$ bonds. An estimate of the value of $\alpha$ was recently obtained by comparing *in sit*u NEXAFS and Raman characterizations of cluster assembled films [22]; by considering the number of bonds for the different hybridizations (2 for $sp$, 3 for $sp^2$) this value can also be used to determine the amount of $sp$ coordinated atoms in the material. By using for $\alpha$ the best estimate from ref. [22] ($\alpha = 1.2$) we obtain that 40% of the atoms of the as deposited film at 150 K are $sp$ hybridized (75% of these contributing to the C2 feature). Remarkably in Fig. 1B the $I_{C1}^{rel}$ and $I_{C2}^{rel}$ behaviors show that only the species producing the C2 line are involved in the overall decrease of the $sp$ population while those producing the C1 line are not affected. These thermally induced modifications appear to be very weak up to 250 K while above this temperature the decay of the C2 peak becomes very fast, suggesting the existence of an activation energy for the decay process.

The C1 contribution has been attributed to polyynes whereas C2 to cumulenes on the basis of simulations [18] and stability evolution against gas exposure [19]. The present results show that we are producing two well distinct *sp* populations with different stabilities and evolutions with temperature, thus supporting the attribution of the C2 species as cumulenes.

Exploiting the selective decay of cumulenes upon heating we have studied their influence on the electrical transport mechanisms of the *sp/sp²* films. In Fig. 2A we plot the logarithm of the current circulating in the films as a function of 1/T for different $I_{C2}^{rel}$ values (i.e. different cumulene contents). To fit these curves we assume a conductivity typical of disordered semiconductors as observed in *sp²* amorphous carbon [1] and in cluster-assembled carbon films [20]. The presence of disorder induces the formation of an electronic density of states characterized by tail bands extending from conduction and valence band edges within a forbidden gap; these tails arise from π states with atypical coordination, while dangling bonds create deep states forming a narrow band close to midgap [1]. Other defects due to impurities or localised structural defects can create discrete energy states or narrow impurity bands within the gap, with acceptor or donor-like behavior [1, 23, 24]. When, as typically is encountered, the energy difference between the Fermi level and the states through which conduction occurs is large compared to thermal energy, the Fermi distribution function can be replaced by a Boltzmann factor and conductivity can be described by:

$$\sigma = \sigma_0 \exp\left(-\frac{E_a}{k_B T}\right) \qquad\qquad (1)$$

where $\sigma_0$ is proportional to the density of states at the conduction or valence band mobility edge, $k_B$ is the Boltzmann constant, and $E_a$ (referred as the activation energy) is the difference between the conduction or valence band mobility edge and the Fermi energy [23]. Since the film geometry and the applied voltage are fixed, $\sigma$ is directly proportional to the current $I$ flowing in the film, and then Equation 1 can be rewritten substituting $\sigma$ and $\sigma_0$ with $I$ and $I_0$ respectively.

The change in the curves shown in Fig. 2A indicates that the electrical transport properties of the $sp/sp^2$ film change during the decay of the cumulenic structures. By performing a linear fit of the Arrhenius plots we found that changes are due to the evolution of both $I_0$ and $E_a$ (Fig. 2B and 2C). In particular, Fig. 2B shows that $I_0$ is strongly affected by the decay of cumulenes, decreasing by more than 80% in the investigated temperature range. The activation energy ($\sim$ 0.16-0.2eV, Fig. 2C) is always definitely lower than the value of 0.3eV as obtained for $sp^2$ cluster-assembled carbon films where cumulenes were absent [20].

These observations are consistent with a model of $sp/sp^2$ carbon consisting of a disordered $sp^2$ network where $sp$ chains are "doping" species, acting as deep defect states pinning the Fermi level out of midgap, and whose density determines $\sigma_0$ (and then $I_0$) like in an extrinsic semiconductor. $Sp^2$ structures can act as stabilizing units capping the cumulene ends thus promoting the stabilizing interaction between the partly delocalized $p$ orbitals of the $sp^2$ amorphous matrix and the conjugated $p$ orbitals of the $sp$ chains [6, 7]. The evolution of $I_0$ with temperature is caused by a decrease in the density of cumulenes. We thus observe a self-doping effect in a pure carbon systems caused by the coexistence of different hybridizations and, in particular, by the metallic character of cumulenes in analogy with self p-doped $sp^3/sp^2$ amorphous carbon [24, 25, 26].

As shown in Fig. 2C, the activation energy $E_a$ decreases of about 40 meV by rising the film temperature and hence decreasing the amount of cumulenes. This can be explained by considering that $sp$ decay takes place through an exothermal cross-linking reaction [4]: the energy released during this process can induce the rearrangement and ordering of a film region surrounding the cumulene chains thus forming graphitic nanodomains whose presence determines the lowering of $E_a$ [1, 26]. The formation of graphitic nanodomains should also modify the Raman spectrum and, in particular, the DG band.

Figure 3A shows the comparison between Raman spectra measured at 150 K and 325 K. The spectra were normalized to their effective intensity $I_{eff}$, defined as $I_{eff}=I_{DG}+\beta \cdot I_C$, where $\beta$ is the ratio between the Raman cross section of $sp^2$ and $sp$ carbon ($1/\alpha$, see ref. [22]) multiplied by the ratio 3/2 between the bond number of the $sp^2$ and $sp$ carbons. By subtracting the two spectra we find that the evolution of the DG band is caused by the appearance of features typical of a disordered graphitic phase characterized by a size of the domains of roughly 1 nm [21]. The formation of graphitic nanoislands in the amorphous $sp^2$ matrix results in a reorganization of the entire film structure with a reduction of the band gap as observed in ion implanted amorphous carbon [26]. In that case the local annealing is caused by the energy deposited in the lattice by the ions, in $sp/sp^2$ carbon the local ordering is promoted by the exothermal cumulene cross-linking reaction.

A further support to this picture is provided by molecular dynamics simulations modelling the evolution of the vibrational density of states (VDOS) of a $sp/sp^2$ carbon network where $sp$ chains undergo cross-linking reactions forming $sp^2$ graphitic structures [27]. Figure 3B shows the VDOS at two different $sp/sp^2$ ratio of the $sp$-$sp$ and $sp^2$-$sp^2$ bond-pairs (circles and open squares in Fig. 3B respectively) obtained as the sum of the autocorrelation function without normalization for systems containing the same number of carbon atoms at a same temperature. We observe a reasonable agreement between measured Raman spectra and calculated VDOS. This comparison is allowed by the strict relationship between the VDOS and the Raman spectrum occurring for highly disordered materials [28]. This suggests that $sp$ evolution as a function of temperature in a $sp^2$ matrix can be modeled as a local graphitization induced by the strong exothermic nature of the cumulene rearrangement reaction.

In conclusion we have presented the production and characterization of a novel $sp/sp^2$ carbon system containing a high (up to 30%) atomic fraction of cumulenic species. This is the first example of a pure carbon system where the $sp$ component influences structural and functional bulk properties.

The presence of cumulene structures affects the electrical transport mechanism of the system through a self-doping mechanism by pinning the Fermi level closer to one of the mobility gap edges. Cumulenes are stable at moderately low temperatures and they decay approaching room temperature causing a local rearrangement of the $sp^2$ structure to form graphitic nanoislands embedded in an amorphous matrix. Our experiments highlight several new aspects relevant for the physics of carbon nanostructures: cumulenic oligomers can be stabilized in a $sp^2$ carbon matrix and used to control the electrical properties and the nanostructure of the material. This system can offer the possibility to study the stabilization mechanisms of $sp$ metallic chains as building blocks of new forms of carbon. By controlling the density and the decay of $sp$ cumulenic species it might be possible to tailor the doping of an amorphous $sp^2$ matrix in view of carbon-based electronic devices [29].

We thank L. Colombo for discussions. This work has been partially supported under project FIRB "Micro and nano carbon structures"

**Figure Captions**

Figure 1 (A) Raman spectra of a cluster-assembled carbon film deposited in UHV conditions on a substrate cooled to 150 K and then heated to 250 K, 275 K, 300 K and 325 K respectively. The two gaussian fit (C1 and C2) of the C band are indicated. (B) Plot of the relative intensity of the C band (circles) as a function of the substrate temperature. The evolution of the relative intensity of the C1 (open triangles) and C2 (open squares) peaks is also shown.

Figure 2 (A) Arrhenius plots of the current flowing in a cluster-assembled film (voltage applied 100 V). Each plot corresponds to one of the Raman spectra shown in Figure 1 and it is then labeled by the relative intensity of the corresponding C2 peak. The linear fits are indicated. (B) Plot of the intercept value of the linear fits ($I_0$) as a function of the relative intensity of the C2 peak ($I_{C2}^{rel}$). (C) Plot of the slope value of the linear fits ($E_a$) as a function of $I_{C2}^{rel}$.

Figure 3 (A) Comparison between Raman spectra of a cluster-assembled film taken at 150 K and 325 K normalized to their effective intensity $I_{eff}$ (see text). The corresponding percentage of hybridised $sp$ atoms is indicated. The difference spectrum, expanded by a factor of two, is indicated at the bottom with the Gaussian fit of D and G bands. (B) Molecular dynamics simulated VDOS of a $sp/sp^2$ system. The two VDOS are calculated for systems containing 48% and 27% of $sp$ hybridized atoms. Their difference is reported at the bottom expanded by a factor of two.

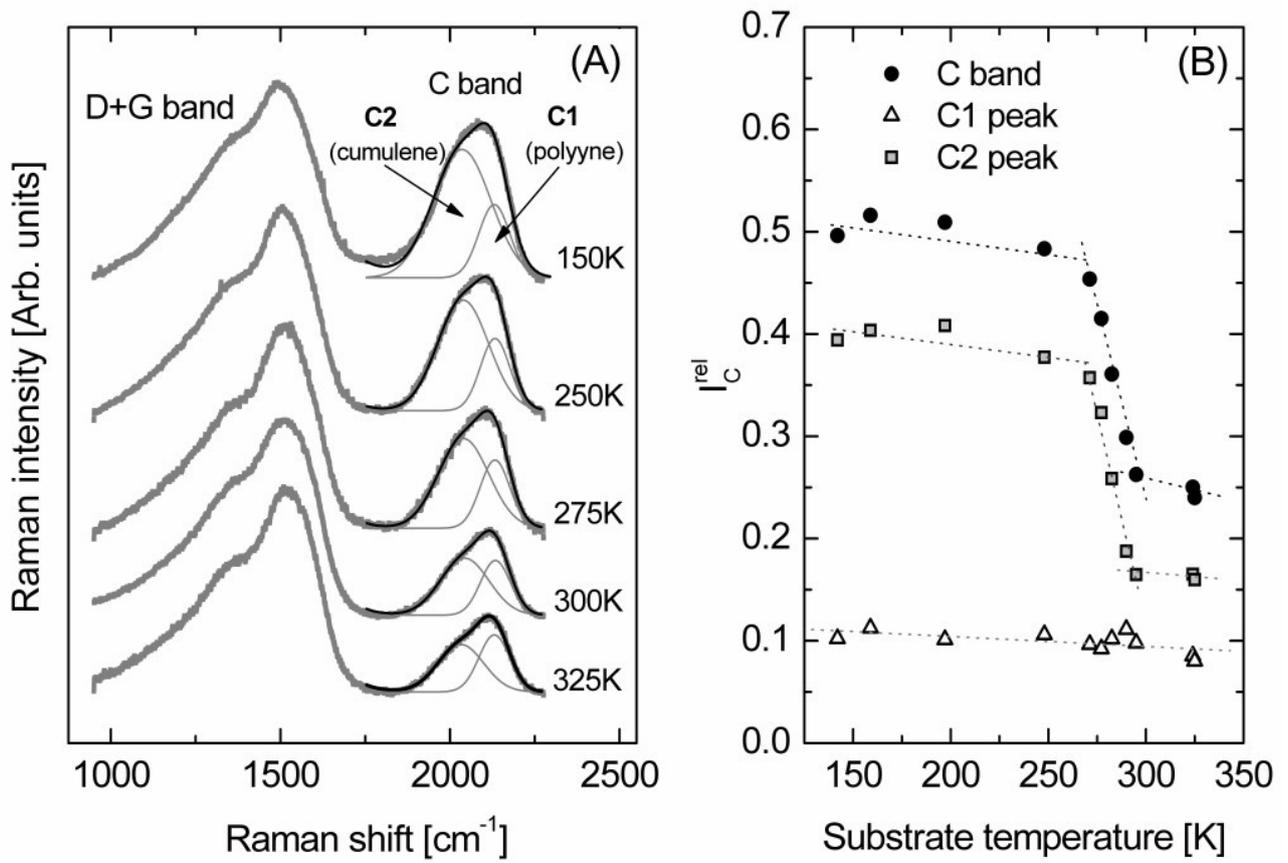

**Figure 1**

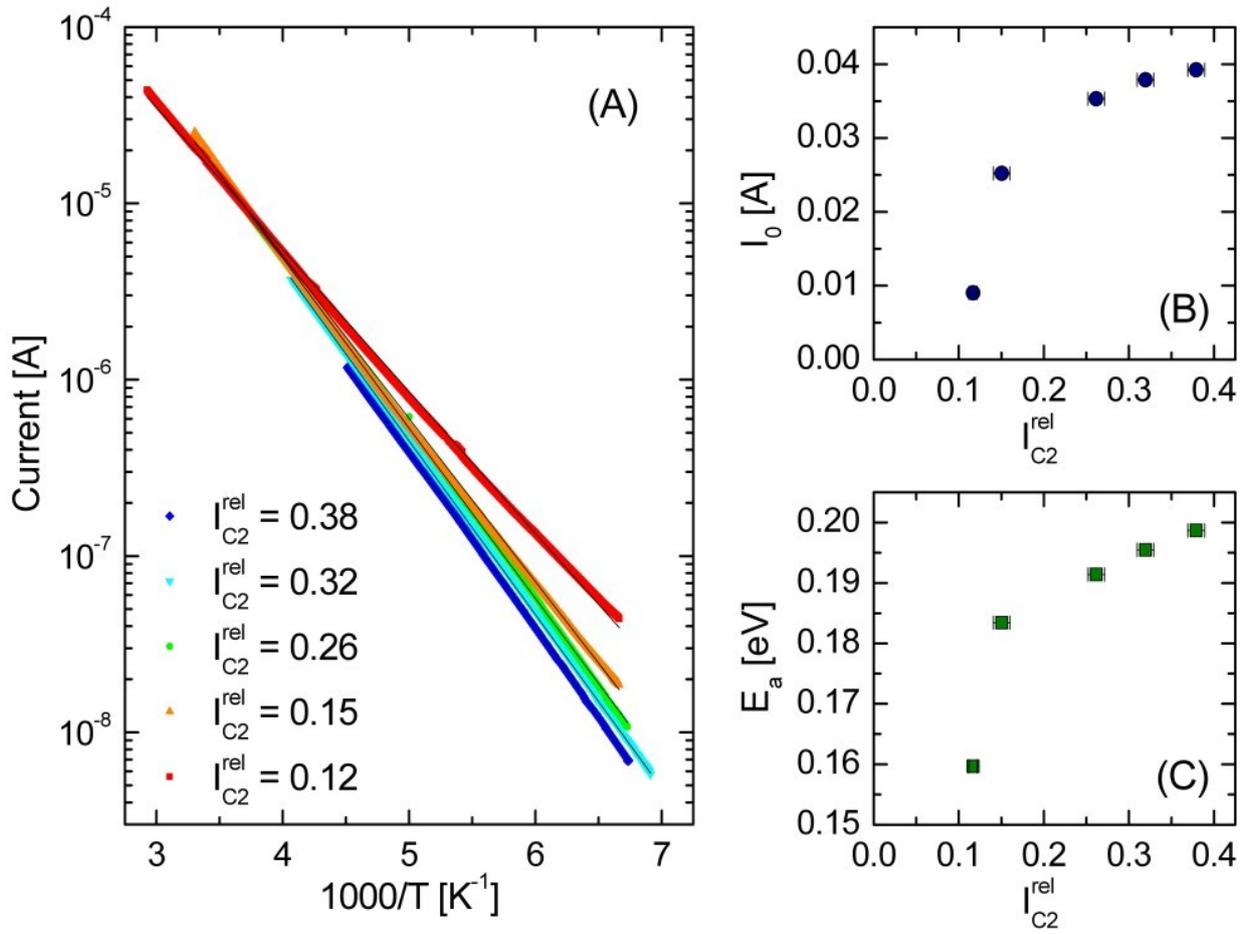

**Figure 2**

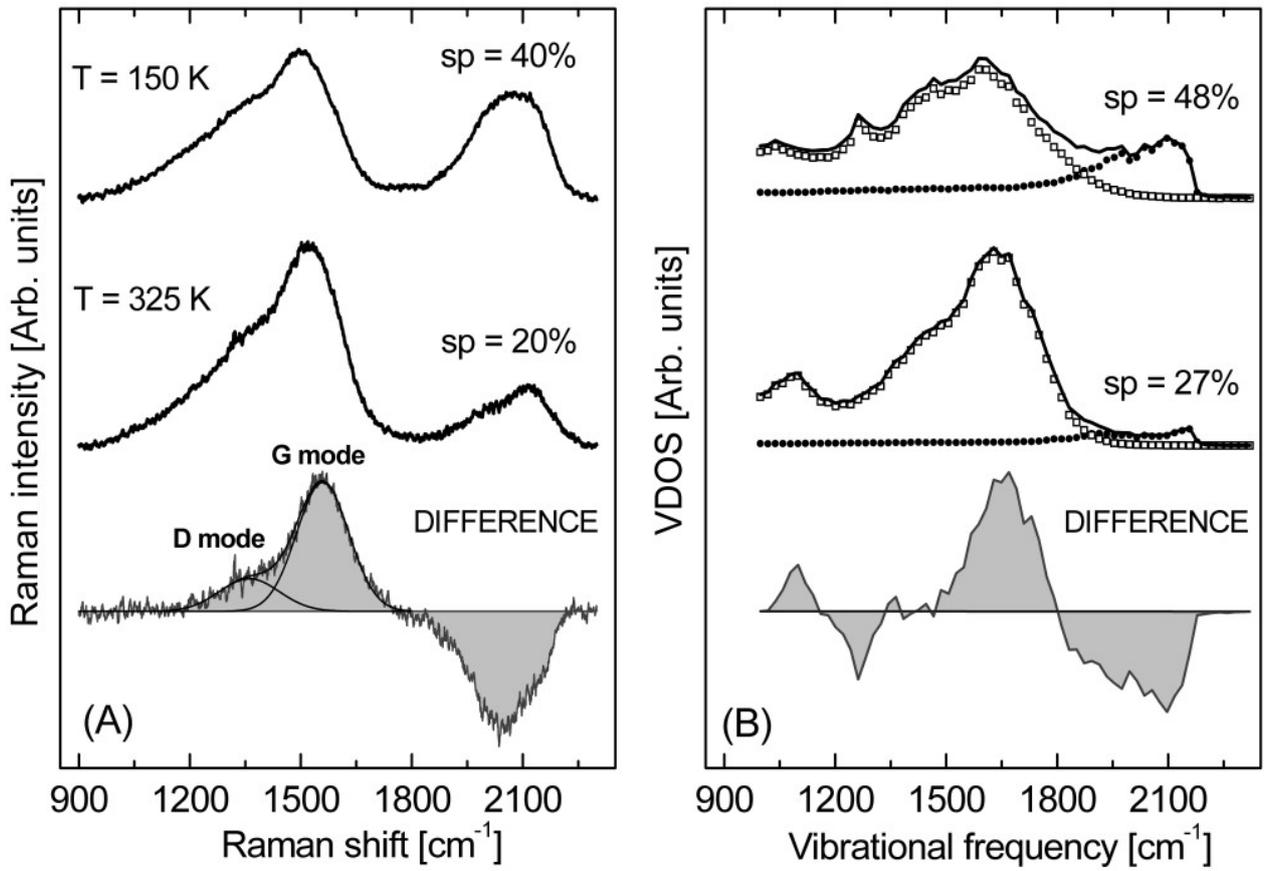

**Figure 3**